\documentclass[aps,prl,twocolumn,groupedaddress]{revtex4-1}
\usepackage{amssymb,amsmath}
\usepackage{color,soul}
\usepackage{graphicx}
\usepackage{hyperref}

\begin{document}

\title{Partial Breaking of Three-Fold Symmetry via Percolation of a Domain Wall}

\author{Soumyadeep Bhattacharya}
\email{sbhtta@imsc.res.in}
\author{Purusattam Ray}
\email{ray@imsc.res.in}
\affiliation{The Institute of Mathematical Sciences,
CIT Campus, Taramani, Chennai 600113, India}

\date{\today}

\begin{abstract}
We show that suppression of vortex strings splits the order-disorder transition
in the three-state Potts ferromagnet on a simple cubic lattice and opens up an
intermediate phase characterized by partial breaking of the three-fold symmetry
and long-range order. In contrast, suppression of vortices in the same model on
a square lattice results in an intermediate phase with enhanced U(1)
symmetry and quasi-long-range order. We show that the difference between the
two phases originates from distinct patterns of domain wall proliferation. A
domain wall, separating the two most populous spin states, percolates on its own
in the former phase but remains at a percolation threshold in the latter.
\end{abstract}


\maketitle

The spontaneous breaking of a three-fold symmetry in three dimensions
is of crucial significance in high energy and condensed matter physics.
The phase transition in the 3+1-dimensional $SU(3)$ gauge theory,
which models the deconfinement of quarks and gluons to a plasma state,
is effectively
described by the transition in a three-dimensional spin model possessing a
three-fold global symmetry~\cite{svetitsky1982critical}. The formation of
cosmic strings across a phase transition during cooling of the early universe
is captured by a model in which the phase of the Higgs field is discretized to three angles~\cite{vachaspati1984formation}.
The three-fold symmetry breaking also captures the behavior of the
valence-bond-solid order parameter across deconfined quantum phase
transitions in the Heisenberg antiferromagnet on a honeycomb lattice~\cite{pujari2013neel,ganesh2013deconfined}.
The phase transitions in these seemingly disparate systems share a
common feature: they are accompanied by the proliferation of vortex defects.

Proliferation of topological defects is the underlying mechanism which
drives phase transitions in a variety of systems possessing
continuous symmetries~\cite{mermin1979topological,chaikin2000principles,vilenkin2000cosmic}.
The superfluid-normal phase transition, in particular,
is driven by the proliferation of vortex defects~\cite{berezinskii1971destruction,kosterlitz1973ordering,kohring1986role,williams1999vortex}.
Does proliferation
of these defects drive the three-fold symmetry breaking transition
as well? The vortex defects need not be solely responsible for driving the
transition.
Recently, we have shown that the three-fold symmetry, which
is broken in the ordered phase of the two-dimensional three-state Potts
model, is restored in the disordered phase by a simultaneous proliferation
of vortices and domain walls~\cite{bhattacharya2016quasi}. When the core energy of vortices in that
model is increased beyond a certain value, the simultaneous proliferation
decouples and the vortices proliferate after the domain walls. This
decoupling splits the order-disorder transition into two and leads to
the appearance of an intermediate phase in which the three-fold symmetry
enhances to U(1). Does the same behavior carry over to the three-dimensional
model as well? Emergence of U(1) symmetry in three dimensions is topic
of debate~\cite{blankschtein1984orderings,miyashita1997nature,oshikawa2000ordered,
lou2007emergence,maes2011rotating,van2011discrete,borisenko2013phase}.
A direct demonstration of such an intermediate phase would
not only help settle the debate but also have fascinating consequences
in the physics of gauge theories and quantum condensed matter systems
effectively described by a three-fold symmetry.

In this paper, we demonstrate that the order-disorder transition in
the three-state Potts ferromagnet on a simple cubic lattice is driven
by the simultaneous proliferation of vortex strings and domain walls.
When we increase the core energy of vortex string segments, the transition
continues to be driven by the coupled proliferation of the defects but
shifts to a higher temperature. Increasing the core energy beyond a certain
value decouples the simultaneous proliferation and splits the transition,
resulting in the formation of an intermediate phase. The intermediate phase
in this case, however, does not exhibit emergence of U(1) symmetry. Instead,
it exhibits partial symmetry breaking. The intermediate phase in both two
and three dimensions results from the proliferation of domain walls. How
can the same defect driven mechanism produce two different types of phases?
In order to find the distinguishing feature between the two proliferation
patterns, we focus on the percolation properties of the domain walls.
We find that one particular type of domain wall percolates on its own in
the intermediate phase of the three-dimensional model while that same
type of domain wall appears to remain at a percolation threshold throughout the
intermediate phase of the two-dimensional model. Our result establishes
that the nature of phases formed by domain wall proliferation can crucially
depend on the percolation behavior of individual types of domain walls.

In order to list the types of domain walls sustained by the model, we place
three-state spins $s_i \in \{0,1,2\}$ at each vertex $i$ of a lattice
$\Lambda$. In this work, we consider the model on a simple cubic lattice and
on a square lattice. The domain walls and vortices
reside on the dual lattice $\Lambda'$, which in the case of the two types of
integer lattices is also an integer lattice, but shifted from $\Lambda$ by
half a lattice spacing along each axis. The domain walls are defects with
codimension one. For the simple cubic lattice, they appear on the plaquettes
of $\Lambda'$ which separate a pair of spins in dissimilar states.
For the square lattice, the domain walls reside on the edges of $\Lambda'$~\cite{bhattacharya2016quasi}.

Vortices are defects with codimension two. On the square lattice, they
reside on the vertices $i' \in \Lambda'$. Each vertex $i'$ is assigned a
winding number $\omega_{i'} = (\Delta_{ba} + \Delta_{cb} + \Delta_{dc} + \Delta_{ad})/3$,
where $\Delta_{ba}$ represents the state difference $(s_b - s_a)$ wrapped
to lie in $[-1,+1]$ and $s_a, s_b, s_c, s_d$ are the spin states
at the four corners of the square plaquette in $\Lambda$ surrounding $i'$.
For the simple cubic lattice, the vortex string segments reside on the
edges $e' \in \Lambda'$ and the winding number $\omega_{e'}$ is calculated
using the same formula but with the spin states at the four corners of
the square plaquette in $\Lambda$ surrounding $e'$~\cite{bittner2005vortex}. The vortex defects
are absent when $\omega = 0$. A vortex or an antivortex is present when
$\omega > 0$ or $\omega < 0$, respectively. The core energy of the vortices
can be increased by associating an energy cost $\lambda$ to each element
of the dual lattice which contains a non-zero winding number~\cite{kohring1986role,
bhattacharya2016quasi,bittner2005vortex,shenoy1990enhancement,sinha2010role}.

\begin{figure}
\includegraphics[width=0.40\textwidth]{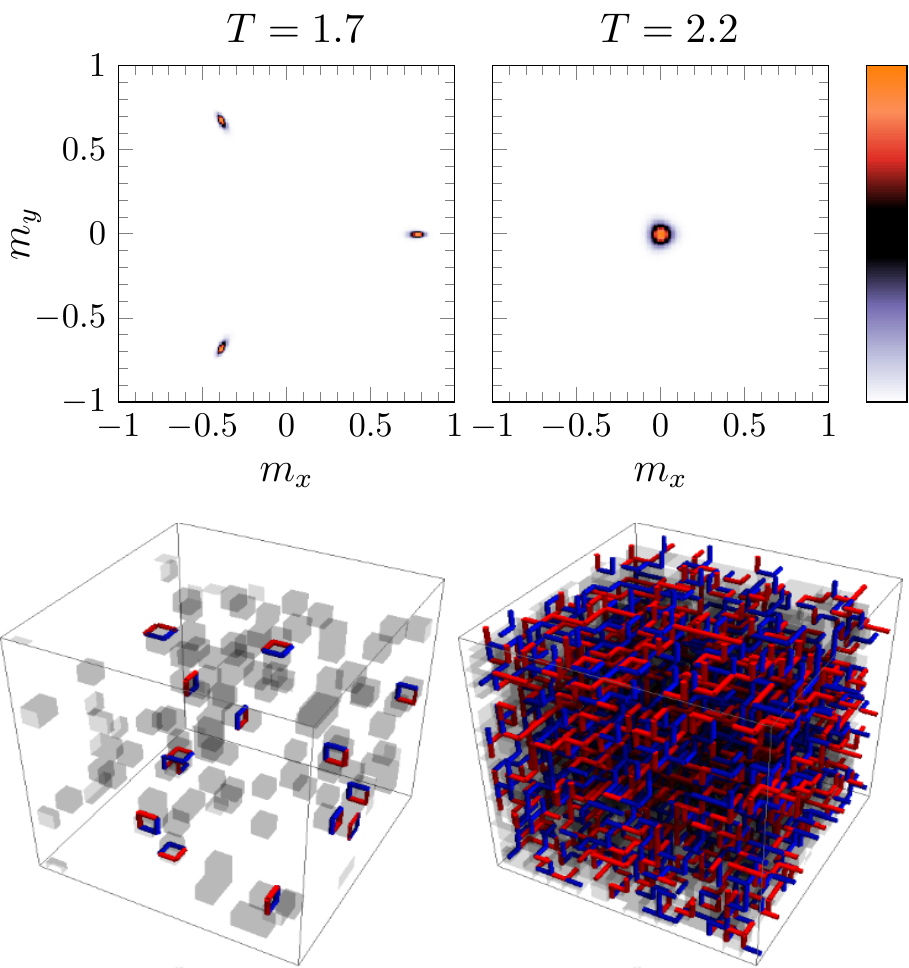}
\caption{Top panel shows the distribution of the three-fold
vector order parameter obtained for a $L=12$ system and bottom
panel shows typical configurations
of domain walls (gray), vortex (blue) and antivortex (red) strings
at two different temperatures: (left) in the ordered phase at $T=1.7$
and, (right) in the disordered phase at $T = 2.2$.}
\label{fig_1}
\end{figure}

The three-state Potts model on the simple cubic lattice with nearest-neighbor
interaction between spins and a $\lambda$ increment of vortex core energy
is described by the Hamiltonian
\begin{eqnarray}
\mathcal{H} = -\sum_{\langle i,j \rangle \in \Lambda} \delta(s_i,s_j)
            + \lambda \sum_{e' \in \Lambda'} |\omega_{e'}|
\label{eqn_hamiltonian}
\end{eqnarray}
We have simulated this model on a lattice with edge length $L$ at
different values of $\lambda$ and temperature $T$ using a single spin-flip
algorithm~\cite{landau2014guide}. At each temperature, we have initialized the system with
a completely ordered spin configuration, discarded the first $10^4$
uncorrelated configurations for equilibriation and measured observables
over the next $10^5$ uncorrelated configurations.
In order to capture the
macroscopic symmetry manifested by the model at different temperatures,
we have measured the three-fold vector order parameter $(m_x,m_y)$, where
$m_x = L^{-3}\sum_{i \in \Lambda} \cos(2\pi s_i/3)$
and $m_y = L^{-3}\sum_{i \in \Lambda} \sin(2\pi s_i/3)$.

In the absence of core energy increment ($\lambda = 0$), the model exhibits
a single order-disorder transition at $T = 1.81$~\cite{janke1997three}. The order parameter
distribution obtained from our simulation clearly shows a breaking of the
three-fold symmetry in the ordered phase and its restoration in the
disordered phase~(Fig.~\ref{fig_1}). In addition, we find that both domain walls and vortex
strings proliferate in the disordered phase while neither of them do so
in the ordered phase.

\begin{figure*}
\includegraphics[width=\textwidth]{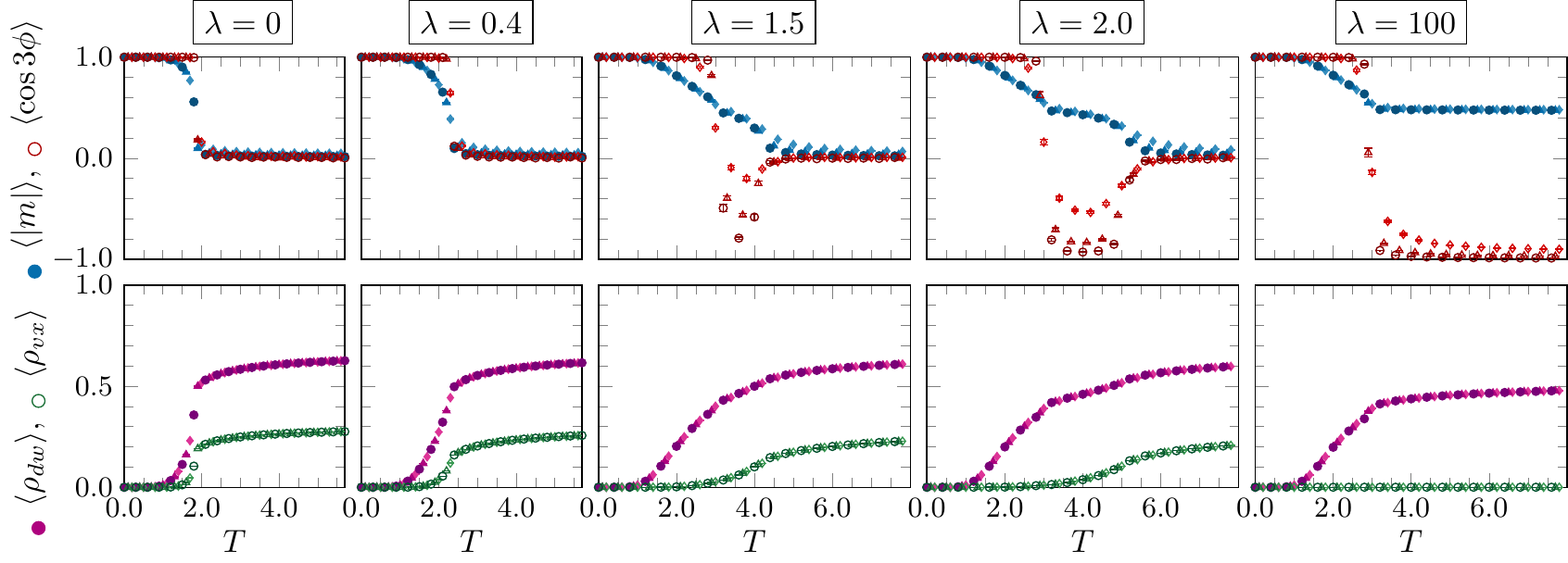}
\caption{The order-disorder transition, in the model without increment in vortex core energy $\lambda$, is
marked by a decay of the magnetization $|m|$ and accompanied by a simultaneous rise in the density of
domain walls $\rho_{dw}$ and the density of vortex strings $\rho_{vx}$. For small values of $\lambda$,
the defect densities rise and the magnetization decays at a higher temperature.
For large values of $\lambda$, the vortex string density begins to rise at a temperature higher than
that of domain wall density. The magnetization shows a two-step decay indicating the appearance of an
intermediate phase. In the intermediate phase, the measure of three-fold symmetry breaking $\langle \cos 3 \phi \rangle$
turns negative.}
\label{fig_2}
\end{figure*}

In order to capture the proliferation behavior in a more quantitative manner,
we have measured the density of the
domain walls $\rho_{dw}$, defined as the fraction of plaquettes
in $\Lambda'$ separating dissimilar spin states, and the density of vortex
strings $\rho_{vx}$, defined as the fraction of edges in $\Lambda'$
that are assigned a non-zero winding number.
For $\lambda = 0$, we find that the densities of both types of defects rise
simultaneously across $T \approx 1.8$~(Fig.~\ref{fig_2}). The corresponding thermodynamic transition
is captured by a decay in the magnetization $|m| = \sqrt{m_x^2 + m_y^2}$ across that
temperature. In order to capture the possibility of symmetry enhancement,
we have measured the strength of three-fold symmetry breaking using
the observable $m_{3 \phi} = \langle \cos 3 \phi \rangle$,
where $\phi = \arctan(m_y/m_x)$ is the angle of the order parameter~\cite{baek2009true}.
The three-fold symmetry is broken in the ordered phase, which is confirmed by
observing that $\langle \cos 3 \phi \rangle = 1$ for $T < 1.8$~(Fig.~\ref{fig_2}).
In the disordered phase both $\langle |m| \rangle$ and $\langle \cos 3 \phi \rangle$
decay to zero.
If the three-fold symmetry enhances to U(1), $\phi$ would fluctuate uniformly between
0 and $2\pi$. This would result in $\langle \cos 3 \phi \rangle = 0$ while the
magnetization remains non-zero.

We begin to gradually increment the core energy in order to
delay the proliferation of the vortex strings and decouple the simultaneous
proliferation. For $\lambda = 0.4$, we find that the density of vortex
strings and domain walls continue to rise together but at a higher temperature
$T = 2.2$~(Fig.~\ref{fig_2}). This forces the order-disorder transition to shift
to a higher temperature, as indicated by the change in the location at which the
magnetization and $\langle \cos 3 \phi \rangle$ decay. We find that the temperature
of simultaneous defect
proliferation and the temperature of the order-disorder transition continues to
shift in this manner upto $\lambda \sim 1.4$. Up till this value, the suppression
of vortex strings is too weak
to decouple the proliferation. Above this value, we being to observe the first
signs of decoupling.

For $\lambda = 1.5$, we find that the vortex string density
rises at a temperature slightly higher than that of the domain walls~(Fig.~\ref{fig_2}). The most
prominent change, however, is visible in the behavior of $\langle \cos 3\phi \rangle$. Across
an intermediate range of temperatures, starting at $T \approx 3$ and ending with
the decay of magnetization at $T = 4.5$, we find that $\langle \cos 3 \phi \rangle$ goes negative.
With increasing $\lambda$, the decay of magnetization, marking the transition
from the intermediate phase to the disordered phase, continues to shift to
higher temperatures following the shift in the proliferation temperature of
vortex strings. For extreme suppression of the vortex strings using
$\lambda = 100$, the intermediate-disorder transition recedes to very high
temperatures and the intermediate region increases in extent. The transition
from the ordered phase to the intermediate phase is accompanied by a rise
in the density of domain walls. This transition remains at $T \approx 3$,
unaffected by the increased suppression of vortex strings.
This result clearly demonstrates that the intermediate-disorder transition
is driven by the proliferation of vortex strings while the order-intermediate
transition is driven by the proliferation of domain walls.
This result is also a source of concern.

\begin{figure}
\includegraphics[width=0.40\textwidth]{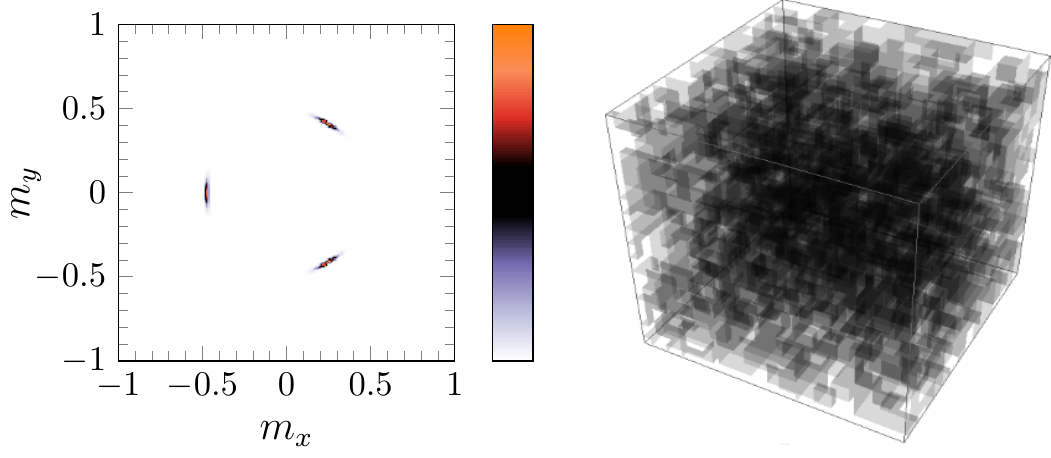}
\caption{Distribution of the order parameter, obtained for a $L=12$ system
in the intermediate phase with $\lambda = 100$ and $T = 6.0$, is shown on the left.
A typical defect configuration obtained at the same temperature (right) shows that
domain walls span across the system while vortex strings are absent. Data has been
obtained for $L=8$ (diamond), $L=12$ (triangle) and $L=16$ (circle).
}
\label{fig_3}
\end{figure}

When the proliferation of vortices in the two-dimensional three-state Potts model
is delayed by raising the core energy of vortices, the order-disorder transition
splits and opens up an intermediate phase in a similar manner~\cite{bhattacharya2016quasi}.
In that case, however, the intermediate phase exhibits enhancement of the three-fold symmetry to U(1).
The intermediate phase in the present model shows a breaking of the three-fold
symmetry at angles $\{\pi/3,\pi,5\pi/3\}$~(Fig.~\ref{fig_3}), which results in
negative values of $\langle \cos 3 \phi \rangle$. In two dimensions, the emergent
U(1) symmetry destroys long-range order in the intermediate phase~\cite{bhattacharya2016quasi}. This forces the
system to quasi-long-range order, due to which the magnetization gradually decays
to zero with increasing system size. In the intermediate phase of the three-dimensional
model, the magnetization remains unchanged with system size but takes up an intermediate value
$\langle |m| \rangle = 0.5$. This suggests that the system is partially ordered
in that phase~(Fig.~\ref{fig_2}). The onset of a partial order is expected because typical
configurations obtained in that phase show that domain walls proliferate~(Fig.~\ref{fig_3}) and allow the system
to fragment into multiple domains which belong to different spin states. However,
the quasi-long-range ordered phase in two dimensions is also formed due to the proliferation
of domain walls~\cite{bhattacharya2016quasi}.
How does the same defect-driven mechanism result in the formation of two different
types of ordered phases?
Since vortex defects are absent in the intermediate phase for both cases~(Fig.~\ref{fig_3}),
it is clear that they do not play a role in determining the nature of the phase.
The distinguishing feature between the phase in the two cases must, therefore, lie
in the pattern of domain wall proliferation alone.
However, the density of domain walls is clearly not a sufficient quantity for
identifying the relevant pattern.

\begin{figure}
\includegraphics[width=0.40\textwidth]{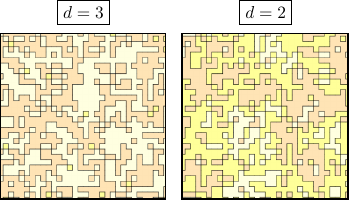}
\caption{Typical spin configurations obtained in the intermediate phase
uncovered by extreme suppression of vortex defects with $\lambda = 100$
for the model on (left) a simple cubic lattice at $T = 6$, for which a
two dimensional slice is shown here, and (right) a square lattice at $T = 4$. Domain
walls are overlaid in black and vortex defects are found to be absent.}
\label{fig_4}
\end{figure}

A visual inspection of typical spin configurations obtained in the intermediate phase
of the two and three-dimensional models reveals a marked difference.
In the configuration of the three-dimensional
model~(Fig.~\ref{fig_4}), we find that the numerous domains mostly belong to
two of the three spin states. This implies that the proliferation of the domain walls
can be further specified as a proliferation of domain walls separating the two spin
states. 
This, however, is not the case for the two-dimensional
model~(Fig.~\ref{fig_4}).
In the configuration obtained for that model, all the three spin states are present.
Although the domain walls appear to span across the system, a particular type of
domain wall, separating any of the two states, appears less likely to span across the
system.
Our inspection suggests that the distinguishing feature in the proliferation pattern
resides in the percolation properties of particular types of domain walls.

We have measured standard percolation observables~\cite{stauffer1994introduction} for
each type of domain wall in the model on both simple cubic and square lattices. A domain wall, which separates a pair
of spin states $a$ and $b$, is assigned a type $(a|b)$.
As the model under considertaion is ferromagnetic, $(a|a)$ does not represent a domain
wall. In addition, $(a|b)$ is equivalent to $(b|a)$, as the interaction between spins
is non-chiral. In order to measure the percolation observables for $(a|b)$ domain
walls, we have joined the $(d-1)$-dimensional $(a|b)$ domain wall segments on the $d$-dimensional dual
lattice $\Lambda'$ only if they share a $(d-2)$-dimensional element of $\Lambda'$.
This criterion specializes to sharing of a dual edge in the three-dimensional case
and the sharing of a dual vertex in the two-dimensional case.
We have distinguished between separate components of $(a|b)$ domain walls by labelling them
using the Hoshen-Kopelman alogrithm~\cite{stauffer1994introduction}.
For each configuration, we have identified the largest domain wall of a particular
type $(a|b)$ and measured the fraction $P_{dw(a|b)}$ of $(d-1)$-dimensional elements
belonging to that domain wall. We have binned the sizes of the remaining $(a|b)$ domain
walls into a distribution $n_{dw(a|b)}(s)$ and calculated the average size of
$(a|b)$ domain walls as~\cite{stauffer1994introduction}
\begin{eqnarray}
\chi_{dw(a|b)} = \frac{\sum_{s} s^2 n_{dw(a|b)}(s)}{\sum_s s n_{dw(a|b)}(s)}
\end{eqnarray}
For each configuration, we have also checked if at least one $(a|b)$ domain wall spans across
the system from a face to its opposite face, under open boundary conditions. The average
of this Boolean measurement over multiple configurations gives the spanning probability
$\Pi_{dw(a|b)}$.
We have also measured these observables for domain walls constructed irrespective
of the particular type and labelled them as $P_{dw}$, $\chi_{dw}$ and $\Pi_{dw}$.

\begin{figure}
\includegraphics[width=0.40\textwidth]{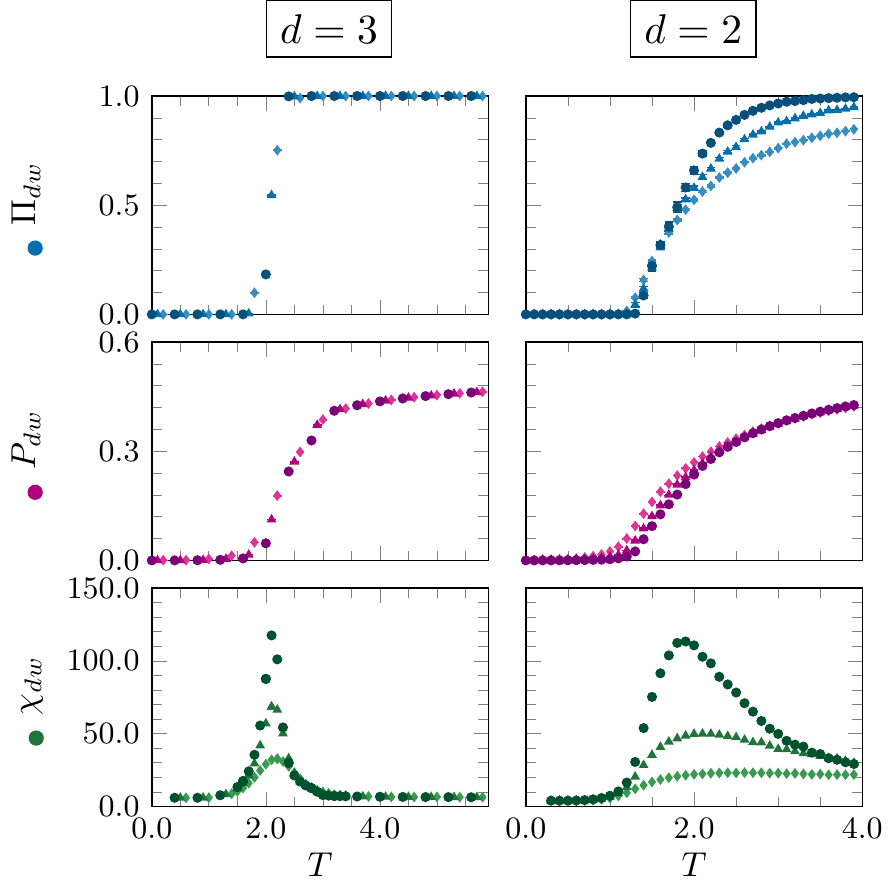}
\caption{Variation of standard percolation observables for domain walls with temperature in
(left) the model on a simple cubic lattice and (right) the model on a square
lattice with $\lambda = 100$. System sizes correspond to those in Fig.~\ref{fig_2}.}
\label{fig_5}
\end{figure}

We find that $\Pi_{dw}$ rises from zero in the ordered phase and saturates to unity
is the intermediate phase of both the two and three-dimensional models~(Fig.~\ref{fig_5}).
In both cases, the percolation
strength $P_{dw}$ rises across the transition from the ordered phase to the intermediate
phase and the average size of domain walls peaks with increasing system size near the
transition in both cases. Although the variation in the percolation observables appear
sharper in the case of the three-dimensional model, they exhibit qualitatively similar
behavior in the two cases. While our result confirms that the domain walls
percolate in the intermediate phase of both models, it suggests that the percolation
properties of the domain walls, irrespective of the particular type, cannot be used to
distinguish between the proliferation pattern in the two cases.

Before presenting the percolation properties for $(a|b)$ domain walls, we mention a
feature of Monte Carlo simulations that needs to be taken into consideration in order
to obtain accurate results for this particular set of observables. All the
observables which we have measured previously are invariant under the symmetry
operations of the $\mathbb{Z}_3$ symmetry group. For example, if all the spins in a
given configuration are rotated by $2\pi/3$, the values of observables like magnetization,
$\langle \cos 3 \phi \rangle$ and even the density of the defects remain invariant.
On the other hand, observables like the percolation strength of
$(0|1)$ domain walls depends on specific states and, therefore, do not remain invariant
under a global rotation of the spins. This becomes a problem in finite size simulations
because the system keeps migrating from one symmetry broken minima to the
other over the course of the simulation~\cite{miyashita1997nature}.
We have mitigated this problem by rotating all the spins by an angle such that the angle
of the symmetry broken minima gets relabelled to 0 (state 0). Since discrete rotations
are constituent members of the $\mathbb{Z}_3$ symmetry group, this procedure of relabelling
keeps the Hamiltonian~(eq. \ref{eqn_hamiltonian}) invariant. However, this relabelling is not sufficient.
Even if we fix the symmetry broken minima, the system can fluctuate
between the angles on the left and right hand side of angle 0 (state 0). In order
to counter such fluctuations, we reflect all the spins in the configuration
across angle 0, in a manner such that the most populous of the two angles
gets relabelled to angle $2\pi/3$ or state $1$. Consequently, the angle on the
other side gets relabelled to $4\pi/3$ or state $2$.
Again, this reflection operation is a constituent member of the $\mathbb{Z}_3$
symmetry group and, therefore, keeps the Hamiltonian~(eq. \ref{eqn_hamiltonian}) invariant.

\begin{figure}
\includegraphics[width=0.40\textwidth]{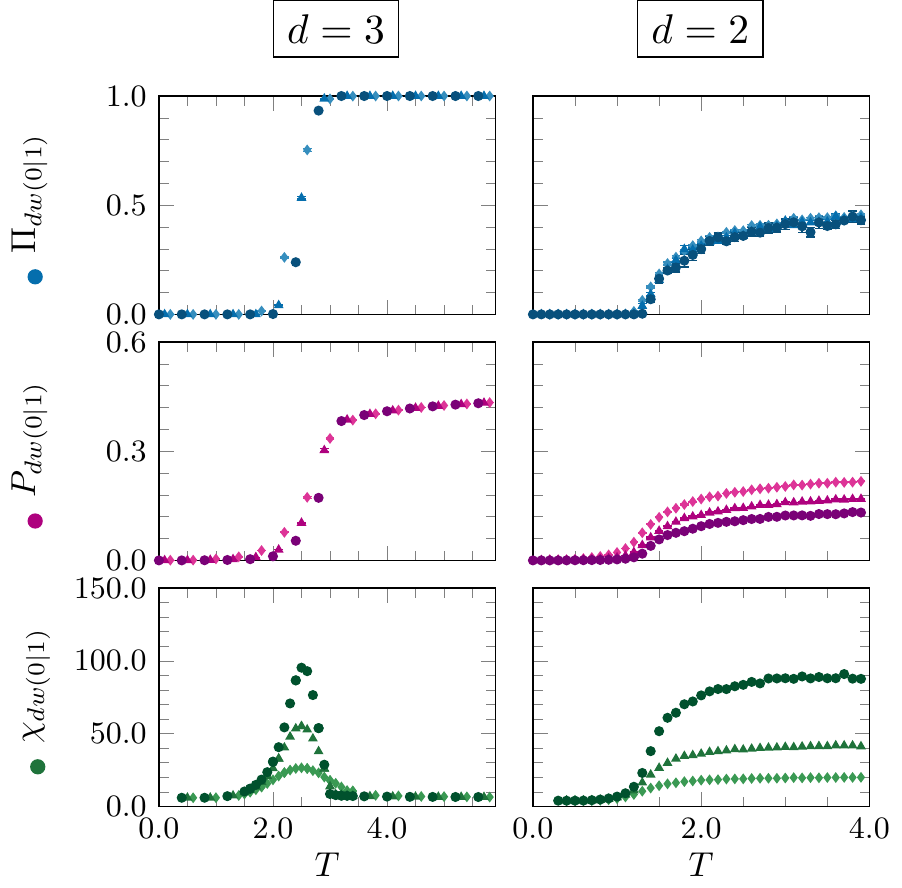}
\caption{Variation of standard percolation observables for $(0|1)$ domain walls with temperature in
(left) the model on a simple cubic lattice and (right) the model on a square
lattice with $\lambda = 100$. System sizes correspond to those in Fig.~\ref{fig_2}.}
\label{fig_6}
\end{figure}

We have applied the relabelling scheme to every configuration generated
in our simulation before measuring the percolation observables for $(a|b)$
domain walls. Under this relabelling, domain walls of type $(0|1)$ separate
spins belonging to the two most populous states.
we find that the $(0|1)$ domain walls begin to
percolate on their own across the transition from the ordered phase to the
intermediate phase in three dimensions~(Fig.~\ref{fig_6}). The spanning probability
of this particular type saturates to unity in the latter phase and the
average size of $(0|1)$ domain walls peaks at the transition.
As expected from the visual inspection~(Fig.~\ref{fig_4}), none of the other types of domain
walls, $(0|2)$ and $(1|2)$, are found to percolate on their own across the transition.

In the intermediate phase of the two-dimensional model, however, we find
that the $(0|1)$ domain walls show a different behavior. The spanning
probability does not saturate to unity but remains at $\Pi_{dw(0|1)} \approx 0.5$~(Fig.~\ref{fig_6}).
The percolation strength of $(0|1)$ domain walls gradually decreases with
increasing system size at each temperature in the phase. In addition,
the average size of $(0|1)$ domain walls not only peaks at the transition
but continues to grow with system size at each temperature in the the phase.
The percolation behavior exhibited at each temperature in the
phase is usually observed only that the threshold of a percolation transition~\cite{stauffer1994introduction}. 
This implies that the $(0|1)$ domain walls do not percolate on their
own in the intermediate phase of the two-dimensional model, but remain
at a percolation threshold throughout the phase. We have found that
none of the other types of domain walls show percolation or threshold
behavior in the intermediate phase.

Our result establishes that the percolation of the $(0|1)$ domain walls is the
salient feature which distinguishes the intermediate phase of the three-dimensional
model from that of its two-dimensional counterpart.
This feature also explains the pattern of symmetry breaking obtained in the
intermediate phase of the three-dimensional model.
The percolation of $(0|1)$ domain walls implies a percolation of
state 0 clusters and state 1 clusters. This, in turn, implies a spin
texture dominated by angles $0$ and $2\pi/3$ in equal proportion. Consequently, the average orientation
of the system becomes $\pi/3$. As the simulation progresses, the symmetry
broken minima shifts across the other two angles as well. Therefore,
we obtain the $\pi/3$ offset pattern of three-fold symmetry breaking in
the ordered parameter distribution for this phase~(Fig.~\ref{fig_3}). Another way to look
at the same result is that state 0 and state 1 act like the
two states of a $\mathbb{Z}_2$ Ising model. In the ordered phase, most
of the spins are in state 0, because of which the Ising symmetry remains
broken. In the intermediate phase, most of the spins arbitrarily pick
up on of the two states. Therefore, the Ising symmetry gets restored,
the symmetry of the system remains only partially broken and the spins
exhibit a partial order instead of complete order. We note that this partial
order is similar to the up-down-up-down height profile of layers grown
in the disordered-flat phase of crystal growth~\cite{weichman1996zippering,ueno1995description}.
This type of partial order has also been reported for $\mathbb{Z}_4$ models
in three dimensions and is termed as a $\langle \sigma \rangle$ phase
in the literature for the Ashkin-Teller model~\cite{ditzian1980phase,pawalicki1997monte}.

The crucial difference between the two-dimensional and three-dimensional
behavior lies in the fact that simultaneous percolation of multiple clusters
(also known as polychromatic percolation) can be sustained by the simple
cubic lattice, due to its higher connectivity, but not by the square lattice~\cite{zallen1977polychromatic}.
Since the two-dimensional system cannot accomodate the simultaneous percolation
of state 0 and state 1 clusters, and yet
the temperature is ripe for domain wall proliferation, the $(0|1)$ domain
walls do not percolate but remain only at a percolation threshold.
While the threshold behavior in two dimensions has not been reported before,
the percolation behavior in three dimensions has been discussed in the
context of a six-state model~\cite{ueno1995description}.

The six-state clock model, with a generalized interaction potential, exhibits a
similar intermediate phase in which the six-fold symmetry is broken at angles offset from
the symmetry breaking pattern in the ordered phase by $\pi/6$~\cite{ueno1993incompletely,todoroki2002ordered}.
This pattern is captured by $\langle \cos 6 \phi \rangle$ going negative in that phase.
It has been suggested
that a variety of intermediate phases in three-dimensional $\mathbb{Z}_n$ models can be distinguished
from each other via the percolation properties of stochastically reduced spin and
bond clusters~\cite{ueno1995description,chayes1997graphical,chayes1998percolation}.
In particular, it has been suggested that the intermediate phase
with negative $\langle \cos n \phi \rangle$ is characterized by the percolation
of a reduced cluster of bonds separating spins which differ by one state~\cite{ueno1995description}.
Such clusters are the stochastically reduced counterparts of the geometric
domain walls that we have considered here. We have shown that a single type of
geometric domain wall percolates on its own in the intermediate phase of the
$\mathbb{Z}_3$ model. It would be interesting to verify if the percolation is
sustained after the stochastic reduction. We note, however, that the suppression
effect produced by the $\lambda$ term in the Hamiltonian~(eq. \ref{eqn_hamiltonian})
would be quite difficult to factor into the scheme for stochastic reduction~\cite{ueno1995description,chayes1997graphical}
as it contains a plaquette-based evaluation of the winding number.


\begin{thebibliography}{}

\bibitem{svetitsky1982critical}
B. Svetitsky and L. G. Yaffe,
Nucl. Phys. B \textbf{210} 423, (1982).

\bibitem{vachaspati1984formation}
T. Vachaspati and A. Vilenkin,
Phys. Rev. D \textbf{30} 2036, (1984).

\bibitem{ganesh2013deconfined}
R. Ganesh R., J. van den Brink and S. Nishimoto,
Phys. Rev. Lett. \textbf{110} 127203, (2013).

\bibitem{pujari2013neel}
S. Pujari, K. Damle and F. Alet,
Phys. Rev. Lett. \textbf{111} 087203 (2013).

\bibitem{mermin1979topological}
N. D. Mermin,
Rev. Mod. Phys. \textbf{51} 591 (1979).

\bibitem{chaikin2000principles}
P. M. Chaikin and T. C. Lubensky
{\it Principles of Condensed Matter Physics}
(Cambridge University Press, 2000).

\bibitem{vilenkin2000cosmic}
A. Vilenkin and E. P. S. Shellard,
{\it Cosmic Strings and Other Topological Defects}
(Cambridge University Press, 2000).

\bibitem{berezinskii1971destruction}
V. L. Berezinskii,
Sov. Phys. JETP \textbf{32} 493 (1971).

\bibitem{kosterlitz1973ordering}
J. M. Kosterlitz and D. J. Thouless,
J. Phys. C \textbf{6} 1181 (1973).

\bibitem{kohring1986role}
G. Kohring, R. E. Shrock and P. Wills,
Phys. Rev. Lett. \textbf{57} 1358 (1986).

\bibitem{williams1999vortex}
G. A. Williams,
Phys. Rev. Lett. \textbf{82} 1201 (1999).

\bibitem{bhattacharya2016quasi}
S. Bhattacharya and P. Ray,
Phys. Rev. Lett. \textbf{116} 097206 (2016).

\bibitem{blankschtein1984orderings}
D. Blankschtein, M. Ma, A. N. Berker, G. S. Grest and C. M. Soukoulis,
Phys. Rev. B \textbf{29} 5250 (1984).

\bibitem{miyashita1997nature}
S. Miyashita
J. Phys. Soc. Jpn. \textbf{66} 3411 (1997).

\bibitem{oshikawa2000ordered}
M. Oshikawa
Phys. Rev. B \textbf{61} 3430 (2000).

\bibitem{lou2007emergence}
J. Lou, A. W. Sandvik and L. Balents
Phys. Rev. Lett. \textbf{99} 207203 (2007).

\bibitem{maes2011rotating}
C. Maes and S. Shlosman,
J. Stat. Phys. \textbf{144} 1238 (2011).

\bibitem{van2011discrete}
A. C. D. Van Enter, C. Kulske and A. A. Opoku,
J. Phys. A \textbf{44} 475002 (2011).

\bibitem{borisenko2013phase}
O. Borisenko, V. Chelnokov, G. Cortese, M. Gravina, A. Papa and I. Surzhikov,
arXiv:1311.0471 [hep-lat] (2013).

\bibitem{bittner2005vortex}
E. Bittner, A. Krinner and W. Janke,
Phys. Rev. B \textbf{72} 094511 (2005).

\bibitem{shenoy1990enhancement}
S. R. Shenoy
Phys. Rev. B \textbf{42} 8595 (1990).

\bibitem{sinha2010role}
S. Sinha and S. K. Roy,
Phys. Rev. E \textbf{81} 041120 (2010).

\bibitem{landau2014guide}
D. P. Landau and K. Binder,
{\it A Guide to Monte Carlo Simulations in Statistical Physics}
(Cambridge University Press, 2014).

\bibitem{janke1997three}
W. Janke and R. Villanova,
Nucl. Phys. B \textbf{489} 679 (1997).

\bibitem{baek2009true}
S. K. Baek, P. Minnhagen and B. J. Kim,
Phys. Rev. E \textbf{80} 060101 (2009).

\bibitem{stauffer1994introduction}
D. Stauffer and A. Aharony,
{\it Introduction to Percolation Theory}
(CRC press, 1994).

\bibitem{weichman1996zippering}
P. B. Weichman and A. Prasad,
Phys. Rev. Lett. \textbf{76} 2322 (1996).

\bibitem{ueno1995description}
Y. Ueno,
J. Stat. Phys. \textbf{80} 841 (1995).

\bibitem{ditzian1980phase}
R. V. Ditzian, J. R. Banavar, G. S. Grest and L. P. Kadanoff,
Phys. Rev. B \textbf{22} 2542 (1980).

\bibitem{pawalicki1997monte}
P. Pawlicki, G. Kamieniarz and L. Debski,
Physica A \textbf{242} 290 (1997).

\bibitem{todoroki2002ordered}
N. Todoroki, Y. Ueno and S. Miyashita,
Phys. Rev. B \textbf{66} 214405 (2002).

\bibitem{ueno1993incompletely}
Y. Ueno and K. Kasono,
Phys. Rev. B \textbf{48} 16471 (1993).

\bibitem{chayes1997graphical}
L. Chayes and J. Machta,
Physica A \textbf{239} 542 (1997).

\bibitem{chayes1998percolation}
L. Chayes, D. McKellar and B. Winn,
J. Phys. A \textbf{31} 9055 (1998).

\bibitem{zallen1977polychromatic}
R. Zallen
Phys. Rev. B \textbf{16} 1426 (1977).

\end{thebibliography}
\end{document}